\documentclass{article}

     \PassOptionsToPackage{numbers, compress}{natbib}

     \usepackage[final]{neurips_2023}

\usepackage[utf8]{inputenc} 
\usepackage[T1]{fontenc}    
\usepackage{setspace}
\usepackage{hyperref}       
\usepackage{url}            
\usepackage{booktabs}       
\usepackage{amsfonts}       
\usepackage{nicefrac}       
\usepackage{microtype}      
\usepackage{xcolor}         

\usepackage{amsmath}
\usepackage{graphicx}
\usepackage{amssymb}

\usepackage{mathtools}
\usepackage{stmaryrd}
\usepackage{algorithm}
\usepackage{algorithmic}

\usepackage{blkarray}
\usepackage{mathdots}
\usepackage{multirow}
\usepackage{numprint}
\usepackage{siunitx}
\usepackage{xcolor, colortbl}
\usepackage{tikz}
\usepackage{wrapfig}
\usepackage{enumitem}
\usepackage{soul}
\usepackage{xurl}

\definecolor{convnext}{RGB}{0,114,178}
\definecolor{slak}{RGB}{204,121,167}
\definecolor{dcls}{RGB}{0,158,115}
\definecolor{slak_sparse}{RGB}{255,193,7}
\definecolor{replknet}{RGB}{213,94,0}
\definecolor{lightcream}{RGB}{255,244,212}
\sethlcolor{lightcream}

\usepackage{multicol}

\definecolor{codegreen}{rgb}{0,0.6,0}
\definecolor{codegray}{rgb}{0.5,0.5,0.5}
\definecolor{codepurple}{rgb}{0.58,0,0.82}
\definecolor{backcolour}{rgb}{0.95,0.95,0.92}

\usepackage{listings}
\lstdefinestyle{mystyle}{
    backgroundcolor=\color{backcolour},   
    commentstyle=\color{codegreen},
    keywordstyle=\color{magenta},
    numberstyle=\tiny\color{codegray},
    stringstyle=\color{codepurple},
    basicstyle=\ttfamily\footnotesize,
    breakatwhitespace=false,         
    breaklines=true,                 
    captionpos=b,                    
    keepspaces=true,                 
    numbers=left,                    
    numbersep=5pt,                  
    showspaces=false,                
    showstringspaces=false,
    showtabs=false,                  
    tabsize=2
}

\title{Audio classification with Dilated Convolution with Learnable Spacings}

\author{%
  Ismail Khalfaoui-Hassani\thanks{Corresponding author} \\
  CerCo UMR 5549, CNRS\\
  Université Toulouse III\\
  Toulouse, France  \\
  \texttt{ismail.khalfaoui-hassani@univ-tlse3.fr} 
  \and
  \textbf{Timothée Masquelier} \\
  CerCo UMR 5549, CNRS\\
  Université Toulouse III\\
  Toulouse, France \\
  \texttt{timothee.masquelier@cnrs.fr} 
  \AND
  \textbf{Thomas Pellegrini} \\
  IRIT, CNRS, Toulouse INP \\
  Université Toulouse III \\
  Toulouse, France \\  
  \texttt{thomas.pellegrini@irit.fr}   
}

\begin{document}

\maketitle

\begin{abstract}
  Dilated convolution with learnable spacings (DCLS) is a recent convolution method in which the positions of the kernel elements are learned throughout training by backpropagation. Its interest has recently been demonstrated in computer vision (ImageNet classification and downstream tasks). Here, we show that DCLS is also useful for audio tagging using the AudioSet classification benchmark. We took two state-of-the-art convolutional architectures using depthwise separable convolutions (DSC), ConvNeXt and ConvFormer, and a hybrid one using attention in addition, FastViT, and drop-in replaced all the DSC layers by DCLS ones. This significantly improved the mean average precision (mAP) with the three architectures without increasing the number of parameters and with only a low cost on the throughput. The method code is based on PyTorch and is available at \href{https://github.com/K-H-Ismail/DCLS-Audio}{https://github.com/K-H-Ismail/DCLS-Audio}.
\end{abstract}

\section{Introduction}

The very popular ConvNeXt model \cite{liu2022convnet}, a fully convolutional model designed for vision tasks, has been successfully adapted to audio classification on AudioSet \cite{pellegrini2023adapting} by transforming audio samples to log-mel spectrograms and adapting the stem of the ConvNeXt model to fit the input audio extracts. This has improved the state of the art of audio classification using convolutional neural networks by achieving better accuracy than PANN-type models \cite{Kong2020}, while having fewer learnable parameters. Furthermore, when used as a backbone for downstream tasks, the ConvNeXt-audio model has achieved positive, if not state-of-the-art, results for the audio captioning and audio retrieval tasks. 

Separately, the Dilated Convolution with Learnable Spacings (DCLS) method has already proven itself in several computer vision tasks \cite{hassani2023dilated}. 
Through a simple drop-in replacement of the model's DSC with DCLS (which can be done automatically for all layers of a model via this script \ref{sec:appendixb}), the DCLS convolution method has empirically proven its effectiveness for several computer vision tasks using ImagNet1k \cite{deng2009imagenet} trained models as backbones. This resulted in a ConvNeXt-dcls model \cite{hassani2023dilated} and a ConvFormer-dcls model \cite{khalfaouihassani2023dilated}, depending on the model chosen in the study, by performing the replacement in the ConvNeXt and ConvFormer models, respectively.

Our aim in the present article is to show empirically that a drop-in replacement of the DCLS method in the same fully convolutional models can improve their accuracy for the task of audio classification on the AudioSet dataset without much effort, demonstrating once again the interest of the method not only on the reference benchmark for image classification but also on the reference benchmark for audio classification (AudioSet \cite{AudioSet}). Furthermore, we add a third test model that differs slightly from the other two in that it's a hybrid model (having both DSC layers and multi-head self-attention layers, depending on the stage to which the layer belongs): FastVit \cite{vasu2023fastvit}, and again, replacing the DSC layers by DCLS improves results. 

This article does not claim to be the absolute state of the art on the task of classification on AudioSet, but rather tries to provide an objective comparison between known and proven convolutional models and those equipped with a DCLS convolution that would make them more efficient.

\section{Related work}

Audio tagging systems were mainly based on convolutional neural networks until recently, with the adaptation of vision transformers to audio processing. The PANN-based models (\textit{e.g.}, CNN14), in particular, comprise blocks of plain $3 \times 3$ kernel convolution layers~\cite{Kong2020}. In~\cite{verbitskiy2022eranns}, PANN-like models were enhanced, in terms of accuracy, model size and inference speed, by adding residual connections, and by modifying the kernel sizes, the stride and padding, using a ``decreasing temporal size parameter''. Other efficient CNN architectures, such as EfficientNet~\cite{gong2021psla}, were also tested in audio tagging. In~\cite{singh2023panns}, efficient PANNs (E-PANN) were obtained by using filter pruning.  In~\cite{drossos2020sound}, DSC layers were used, which resulted in large reductions in model complexity, together with performance gains. In~\cite{pellegrini2023adapting}, doing so in PANN's CNN14 also yielded significant model size reduction (about 60\% relative), whilst observing a gain in performance. In this last study, ConvNeXt was adapted to perform the audio tagging task in AudioSet. It performed better or on par with the transformer-based architectures AST~\cite{gong2021ast} and PaSST-S~\cite{koutini2021efficient}.

\section{Methods}

\subsection{Dataset and configuration}
\textbf{Dataset.} 
In all the experiments in this article, we used AudioSet \cite{AudioSet}, the reference dataset in audio classification. It contains about 2 million video clips downloaded from the YouTube platform. We are only interested in the audio portion of these clips and are not using the video dataset. The audio clips available in AudioSet can vary in size, but most are 10 seconds long. If a sample is longer than that, we truncate it; if shorter, we pad it with zeros. The classification task in AudioSet consists of assigning each sample to the class or classes to which it belongs among the 527 available labels. It is thus a multi-label classification task. The majority of the excerpts correspond to one of the two classes "speech" and "music" (often both), due to their predominance on the aforementioned video hosting site. This latter fact leads to an imbalance in the dataset, with several classes being poorly represented while a few classes account for most of the dataset. 
We downloaded the data in 2018, and some of the YouTube links have been broken since then. Our AudioSet data contains 1,921,982 clips (unbalanced train), 21,022 clips (balanced train), and 19,393 clips (evaluation).

\textbf{Metrics.} \label{sec:metrics} We report the usual evaluation metric for AudioSet tagging: mean average precision (mAP) 
which is typically the metric of interest in audio tagging. 
All DCLS-equipped models studied here outperform their respective baselines using this metric.

\textbf{No weighted sampler.} Given the unbalanced nature of the dataset, many state-of-the-art models make good use of a weighted random sampler \cite{Kong2020, gong2022contrastive, huang2022mavil}, where each class in the dataset is weighted by its frequency of occurrence in the dataset. This is a classic machine learning approach to mitigate data imbalance. However, as pointed out by \cite{moore2023dataset}, these approaches based on a weighted sampler whose oversampling rate is adjusted as a training hyperparameter seem to overfit the dataset more than anything else and do not favor the rarest classes. Since in this article, we are only interested in the comparative study between baseline models and the same models augmented with the DCLS method, we have chosen not to include weighted samplers in our training phases, even if this means losing a few points in mAP, thereby allowing a comparison that is less noisy due to the effects of sampling. Furthermore, the naive use of Mixup augmentation \cite{zhang2018mixup} in conjunction with a weighted sampler may turn out to be a source of undesirable behavior, since proceeding in this way could destroy the weighted sampling that was originally intended, as the Mixup acts randomly by drawing two samples without taking balancing into account. Weighting-aware approaches to Mixup such as \cite{Ramasubramanian2023selmix} should be better investigated and implemented in order to better take advantage of both methods.

\textbf{Spectrogram resolution.} Many audio classification models use raw audio signals \cite{oord2016wavenet, dai2017very, akbari2021vatt}, while a growing number of state-of-the-art models use spectrograms, taking advantage of the signal's periodic aspect by using the Short-time Fourier transform \cite{Kong2020, koutini2021efficient, huang2022mavil}. We prefer this second choice in order to use computer vision baselines. Additionally, the obtained spectrograms are often filtered using the mel psychoacoustic scale \cite{stevens1937scale}. We use the latter filtering to obtain mel-frequency spectrograms, which we transform from the power/amplitude scale to the decibel scale. A comprehensive enumeration of the hyperparameters used to perform these transformations is given in Table~\ref{tab:2}.

The final spectrogram size obtained is ($F=128$, $T=1001$). In the course of our experiments, we noticed that the larger the size of the spectrograms (in both frequency and time), the greater the mAP of the models, but to the detriment of their throughput. This is a well-known phenomenon in computer vision, where higher input image resolution often leads to better vision model accuracy but also to higher computational time costs. We believe that this resolution provides a good trade-off in mAP-throughput and argue that there is a sweet spot between resolution and stem size that offers optimal performance regarding mAP-throughput.

\textbf{Adapting the stem.} The three neural networks used in this study all come from the world of computer vision. It is therefore essential to adapt the stem of these models in order to process no longer natural images made up of three channels corresponding to the RGB colors, but instead a spectrogram with a single channel and a size different from the crop images initially designed for vision tasks. To this end, we used a basic stem, common to all three studied models, namely a convolution layer with a kernel size = (2, 16) and a stride = (2, 16). This stem produces maps of size (64, 62) from input spectrograms of size (128, 1001). This type of stem is similar to that originally found in the ConvNeXt model, while the ConvFormer model featured a slightly more sophisticated stem with a kernel size larger than the stride size. The FastVit model, on the other hand, came with a much more complex stem consisting of several layers based on the MobileOne block \cite{vasu2023mobileone}. 

Imposing a common stem on all the models in the study means that, on the one hand, we can compare the models more accurately, knowing that the input resolution will be the same for all of them. On the other hand, in the absence of an optimal stem adapted to audio spectrograms, we use a coarse stem with which we can conduct our study. Note that the search for an ideal stem is a study in itself and that the stem presented here can always be refined and improved.

\textbf{Pretraining on ImageNet1k.} Using pre-trained models on ImageNet \cite{deng2009imagenet} as a better initialization to solve the tagging task on AudioSet is common practice~\cite{gong2021ast,koutini2021efficient,pellegrini2023adapting}. In the first few epochs, models initialized in this way have a clear advantage over those initialized randomly. However, this advantage is quickly regained over the course of training, and randomly initialized models often end up performing similarly to or slightly worse than pre-trained ones. We only use pre-trained models on ImageNet1k when they are available and do not cost us anything to train. Therefore, we use the symbol $\ddag$ to designate models that have not been pre-trained on ImageNet.

\textbf{Configuration.} In Table~\ref{tab:2} is a comprehensive list of the hyperparameters and augmentations used in this study. These are largely similar to the hyperparameters used in \cite{liu2022convnet}. Note that a high drop path (0.4) is used in this work to overcome the overfitting problem encountered with the tagging task on AudioSet and that large effective batch sizes were used (4096) to speed up training. However, some instabilities during training were noted, particularly for the ConvFormer model. These instabilities are known from \cite{yu2022metaformer} and were resolved by using the LAMB optimizer \cite{you2019large}, while we used AdamW \cite{adamw} for the other two models. 
\vspace{-0.2cm}
\subsection{Models}

\label{sec:models}
We used three different models from computer vision that we adapted to audio inputs to corroborate our results. The first one is a fully convolutional model: ConvNeXt-tiny \cite{liu2022convnet}. The second one is also a purely convolutional model that outperforms the ConvNext model on the ImageNet classification task: ConvFormer-S18 \cite{yu2022metaformer}. The DCLS method as a replacement for DSC has already been successfully used in the latter two models for various vision tasks, including image classification on ImageNet1k \cite{hassani2023dilated, khalfaouihassani2023dilated}. The third model we used is more recent and achieves the current state-of-the-art throughput-accuracy trade-off in image classification on the ImageNet dataset: FastVit-SA24 \cite{vasu2023fastvit}. The latter is a so-called hybrid model, i.e., it contains DSC layers as well as and multi-head self-attention layers. 

\subsection{DCLS substitution}

Considering the baseline models discussed in the previous section, we carried out the following study: we trained the baselines on the concatenation of the unbalanced train and the balanced train sets of AudioSet, then evaluated them on the evaluation subset. We repeated the same process with the same models, except that this time we replaced all DSC layers having a kernel size equal to 7 with a DCLS convolution layer. In all test cases, we used exactly the same training configuration to avoid attributing performance gains to any reason other than the replacement of the DSC layers by DCLS ones. Also, to learn the positions (and standard deviations for DCLS-Gauss) of each kernel element, we followed the same training techniques as those listed in \cite{khalfaouihassani2023dilated}. This gave us 6 test cases to examine in total, for which we measure the mAP metric mentioned in Section~\ref{sec:metrics} averaged over three different seeds (seeds 0, 1, and 2). 

\section{Results}
The results presented in Table~\ref{tab:1} demonstrate the performance of the three models mentioned in Section \ref{sec:models} on the $128 \times 1001$ spectrograms, where the convolution method used varies. Notably, we observe that 
when comparing each baseline model with its DCLS-equipped counterpart, the use of DCLS-Gauss with a kernel size of $23^2$ and a kernel count of $26$ stands out, achieving a higher mAP (+0.6 on average) with an equal or lower number of parameters. This result highlights the effectiveness of DCLS-Gauss in enhancing classification performance. 
DCLS does, however, introduce a reduction in throughput ($13\%$ for ConvNeXt-T and FastVit-SA24 and $23\%$ for ConvFormer-S18) due to the use of larger kernels. The results of a previous study on ConvNeXt~\cite{pellegrini2023adapting} show that an mAP of $47.1$ can be achieved, but here we only reach $44.8$ for the baseline; this is due to the fact that in that previous study, a higher spectrogram resolution was used ($224 \times 1001$ versus $128 \times 1001$ in this work) and that a stem size of $4 \times 4$ instead of $2 \times 16$ here was used to produce larger feature maps, which is reflected both in the large memory required to run this model and in the model's throughput.

\begin{table*}[!htbp]
\begin{center}
\resizebox{\textwidth}{!}{
$
\begin{array}{lccccc}
\toprule
\text {model}
&\begin{array}{l}
\text { ker. size } \\
\text { / count  }
\end{array} & \text {method}  & \text { \# param.} & \text { mAP }  & \begin{array}{l}
\text { throughput } \\
\text { (sample / s) }
\end{array} \\
\hline
\text {CNN14 \cite{Kong2020}} &  & \text{Conv.} & 80.7\mathrm{M} & 43.1 & 378.2 \\
\text {PaSST-S \cite{koutini2021efficient}} &  & \text{MHS. Attention.} & 87\mathrm{M} & 47.1 & 88.7 \\
\text {ConvNeXt-T \cite{pellegrini2023adapting}} & 7^2 \ / \ 49 & \text{Depth. Conv.} & 28.2\mathrm{M} & 47.1 & 153.6 \\
\hline 
\text {ConvFormer-S18}^\dag & 7^2 \ / \ 49 & \text{Depth. Conv.} & 26.8 \mathrm{M} & 43.14 \pm 0.03 & 513.3 \\
\rowcolor{lightcream}\text {ConvFormer-S18}^\dag & 23^2  \ / \  26 & \text{DCLS-Gauss} & 26.8 \mathrm{M} & 43.68 \pm 0.02 & 396.8 \\

\text {FastVIT-SA24}^\ddag & 7^2 \ / \ 49 & \text{Depth. Conv.} & \textbf{21.5} \mathrm{M} & 43.82 \pm 0.05 & \textbf{633.6}\\
\rowcolor{lightcream}\text {FastVIT-SA24}^\ddag  & 23^2  \ / \  26 & \text{DCLS-Gauss} & \textbf{21.5} \mathrm{M} & 44.4 \pm 0.07 & 551.7 \\

\text {ConvNeXt-T } & 7^2 \ / \ 49 & \text{Depth. Conv.} & 28.6 \mathrm{M} & 	44.83 \pm 0.14  & 591.4 \\
\rowcolor{lightcream}\text {ConvNeXt-T } & 23^2  \ / \  26 & \text{DCLS-Gauss} & 28.6 \mathrm{M} & \textbf{45.52} \pm 0.05  & 509.4 
\end{array}
$
}
\end{center}
\caption{\textbf{Classification mean average precision (mAP) on the evaluation set of AudioSet.} For the baselines using DSC and the DCLS-Gaussian cases, the results have been averaged over 3 distinct seeds and presented in the format mean $\pm$ standard deviation. $\dag:$ trained using LAMB, $\ddag:$ no ImageNet pretraining. The throughputs were calculated with a single NIVIDIA V100 32-GB gpu.}
\label{tab:1}
\end{table*}

\section{Conclusion}
In conclusion, this article has demonstrated the efficacy of Dilated Convolution with Learnable Spacings (DCLS) as a method with promising applications beyond the computer vision field. By exploiting DCLS in the audio tagging task on AudioSet, we have demonstrated tangible improvements in accuracy when compared to models employing traditional DSC methods. While this work does not claim to establish an absolute state-of-the-art benchmark, it does contribute valuable insights into the potential of DCLS convolution in audio classification. This research underscores the significance of exploring novel convolutional techniques, like DCLS, and adapting them to various domains beyond their initial design. As the field of deep learning continues to evolve, such methods pave the way for broader and more efficient applications, thereby advancing the state of the art in deep learning.


\section*{Acknowledgments}
This work was performed using HPC resources from GENCI–IDRIS (Grant 2023-[AD011013219R1]). Support from the ANR-3IA Artificial and Natural Intelligence Toulouse Institute is gratefully acknowledged. We would also like to thank the region of Toulouse Occitanie.

\bibliographystyle{abbrv}
\bibliography{references}

\begin{thebibliography}{10}

\bibitem{akbari2021vatt}
H.~Akbari, L.~Yuan, R.~Qian, W.-H. Chuang, S.-F. Chang, Y.~Cui, and B.~Gong.
\newblock Vatt: Transformers for multimodal self-supervised learning from raw
  video, audio and text.
\newblock {\em Advances in Neural Information Processing Systems},
  34:24206--24221, 2021.

\bibitem{dai2017very}
W.~Dai, C.~Dai, S.~Qu, J.~Li, and S.~Das.
\newblock Very deep convolutional neural networks for raw waveforms.
\newblock In {\em 2017 IEEE international conference on acoustics, speech and
  signal processing (ICASSP)}, pages 421--425. IEEE, 2017.

\bibitem{deng2009imagenet}
J.~Deng, W.~Dong, R.~Socher, L.-J. Li, K.~Li, and L.~Fei-Fei.
\newblock Imagenet: A large-scale hierarchical image database.
\newblock In {\em Proc. IEEE/CVF Conf. Comput. Vis. Pattern Recog. (CVPR)},
  pages 248--255. IEEE, 2009.

\bibitem{drossos2020sound}
K.~Drossos, S.~I. Mimilakis, S.~Gharib, Y.~Li, and T.~Virtanen.
\newblock Sound event detection with depthwise separable and dilated
  convolutions.
\newblock In {\em 2020 International Joint Conference on Neural Networks
  (IJCNN)}, pages 1--7. IEEE, 2020.

\bibitem{AudioSet}
J.~F. Gemmeke, D.~P.~W. Ellis, D.~Freedman, A.~Jansen, W.~Lawrence, R.~C.
  Moore, M.~Plakal, and M.~Ritter.
\newblock Audio set: An ontology and human-labeled dataset for audio events.
\newblock In {\em Proc. IEEE ICASSP}, New Orleans, LA, 2017.

\bibitem{gong2021ast}
Y.~Gong, Y.-A. Chung, and J.~Glass.
\newblock {AST: Audio Spectrogram Transformer}.
\newblock In {\em Proc. Interspeech}, pages 571--575, Brno, 2021.

\bibitem{gong2021psla}
Y.~Gong, Y.-A. Chung, and J.~Glass.
\newblock {PSLA: Improving audio tagging with pretraining, sampling, labeling,
  and aggregation}.
\newblock {\em IEEE/ACM Transactions on Audio, Speech, and Language
  Processing}, 29:3292--3306, 2021.

\bibitem{gong2022contrastive}
Y.~Gong, A.~Rouditchenko, A.~H. Liu, D.~Harwath, L.~Karlinsky, H.~Kuehne, and
  J.~R. Glass.
\newblock Contrastive audio-visual masked autoencoder.
\newblock In {\em The Eleventh International Conference on Learning
  Representations}, 2022.

\bibitem{huang2022mavil}
P.-Y. Huang, V.~Sharma, H.~Xu, C.~Ryali, H.~Fan, Y.~Li, S.-W. Li, G.~Ghosh,
  J.~Malik, and C.~Feichtenhofer.
\newblock Mavil: Masked audio-video learners.
\newblock {\em arXiv preprint arXiv:2212.08071}, 2022.

\bibitem{hassani2023dilated}
I.~Khalfaoui-Hassani, T.~Pellegrini, and T.~Masquelier.
\newblock Dilated convolution with learnable spacings.
\newblock In {\em The Eleventh International Conference on Learning
  Representations}, 2023.

\bibitem{khalfaouihassani2023dilated}
I.~Khalfaoui-Hassani, T.~Pellegrini, and T.~Masquelier.
\newblock Dilated convolution with learnable spacings: beyond bilinear
  interpolation.
\newblock In {\em ICML 2023 Workshop on Differentiable Almost Everything:
  Differentiable Relaxations, Algorithms, Operators, and Simulators}, 2023.

\bibitem{ko15_interspeech}
T.~Ko, V.~Peddinti, D.~Povey, and S.~Khudanpur.
\newblock {Audio augmentation for speech recognition}.
\newblock In {\em Proc. Interspeech 2015}, pages 3586--3589, 2015.

\bibitem{Kong2020}
Q.~Kong, Y.~Cao, T.~Iqbal, Y.~Wang, W.~Wang, and M.~D. Plumbley.
\newblock Panns: Large-scale pretrained audio neural networks for audio pattern
  recognition.
\newblock {\em IEEE/ACM Transactions on Audio, Speech, and Language
  Processing}, 28:2880--2894, 2020.

\bibitem{koutini2021efficient}
K.~Koutini, J.~Schl{\"u}ter, H.~Eghbal-zadeh, and G.~Widmer.
\newblock Efficient training of audio transformers with patchout.
\newblock In {\em Proc. Interspeech}, pages 2753--2757, Incheon, 2022.

\bibitem{larsson2016fractalnet}
G.~Larsson, M.~Maire, and G.~Shakhnarovich.
\newblock Fractalnet: Ultra-deep neural networks without residuals.
\newblock In {\em International Conference on Learning Representations}, 2016.

\bibitem{liu2022convnet}
Z.~Liu, H.~Mao, C.-Y. Wu, C.~Feichtenhofer, T.~Darrell, and S.~Xie.
\newblock A convnet for the 2020s.
\newblock In {\em Proc. IEEE/CVF Conf. Comput. Vis. Pattern Recog. (CVPR)},
  pages 11976--11986, 2022.

\bibitem{loshchilov2016sgdr}
I.~Loshchilov and F.~Hutter.
\newblock Sgdr: Stochastic gradient descent with warm restarts.
\newblock In {\em International Conference on Learning Representations}, 2016.

\bibitem{adamw}
I.~Loshchilov and F.~Hutter.
\newblock Decoupled weight decay regularization.
\newblock In {\em International Conference on Learning Representations}, 2017.

\bibitem{moore2023dataset}
R.~C. Moore, D.~P. Ellis, E.~Fonseca, S.~Hershey, A.~Jansen, and M.~Plakal.
\newblock Dataset balancing can hurt model performance.
\newblock In {\em ICASSP 2023-2023 IEEE International Conference on Acoustics,
  Speech and Signal Processing (ICASSP)}, pages 1--5. IEEE, 2023.

\bibitem{oord2016wavenet}
A.~v.~d. Oord, S.~Dieleman, H.~Zen, K.~Simonyan, O.~Vinyals, A.~Graves,
  N.~Kalchbrenner, A.~Senior, and K.~Kavukcuoglu.
\newblock {WaveNet: A generative model for raw audio}.
\newblock {\em arXiv preprint arXiv:1609.03499}, 2016.

\bibitem{pellegrini2023adapting}
T.~Pellegrini, I.~Khalfaoui-Hassani, E.~Labbé, and T.~Masquelier.
\newblock {Adapting a ConvNeXt Model to Audio Classification on AudioSet}.
\newblock In {\em Proc. INTERSPEECH 2023}, pages 4169--4173, 2023.

\bibitem{Ramasubramanian2023selmix}
S.~Ramasubramanian, H.~Rangwani, S.~Takemori, K.~Samanta, Y.~Umeda, and
  R.~Venkatesh~Babu.
\newblock Selmix: Selective mixup fine tuning for optimizing non-decomposable
  metrics.
\newblock In {\em The Differentiable Almost Everything Workshop of the 40 th
  International Conference on Machine Learning}, 2023.

\bibitem{singh2023panns}
A.~Singh, H.~Liu, and M.~D. Plumbley.
\newblock {E-PANNs: Sound Recognition Using Efficient Pre-trained Audio Neural
  Networks}.
\newblock In {\em Proc. Inter Noise}, Chiba, 2023.

\bibitem{stevens1937scale}
S.~S. Stevens, J.~Volkmann, and E.~B. Newman.
\newblock A scale for the measurement of the psychological magnitude pitch.
\newblock {\em The journal of the acoustical society of america},
  8(3):185--190, 1937.

\bibitem{szegedy2016rethinking}
C.~Szegedy, V.~Vanhoucke, S.~Ioffe, J.~Shlens, and Z.~Wojna.
\newblock Rethinking the inception architecture for computer vision.
\newblock In {\em Proceedings of the IEEE conference on computer vision and
  pattern recognition}, pages 2818--2826, 2016.

\bibitem{vasu2023fastvit}
P.~K.~A. Vasu, J.~Gabriel, J.~Zhu, O.~Tuzel, and A.~Ranjan.
\newblock Fastvit: A fast hybrid vision transformer using structural
  reparameterization.
\newblock {\em arXiv preprint arXiv:2303.14189}, 2023.

\bibitem{vasu2023mobileone}
P.~K.~A. Vasu, J.~Gabriel, J.~Zhu, O.~Tuzel, and A.~Ranjan.
\newblock Mobileone: An improved one millisecond mobile backbone.
\newblock In {\em Proceedings of the IEEE/CVF Conference on Computer Vision and
  Pattern Recognition}, pages 7907--7917, 2023.

\bibitem{verbitskiy2022eranns}
S.~Verbitskiy, V.~Berikov, and V.~Vyshegorodtsev.
\newblock Eranns: Efficient residual audio neural networks for audio pattern
  recognition.
\newblock {\em Pattern Recognition Letters}, 161:38--44, 2022.

\bibitem{you2019large}
Y.~You, J.~Li, S.~Reddi, J.~Hseu, S.~Kumar, S.~Bhojanapalli, X.~Song,
  J.~Demmel, K.~Keutzer, and C.-J. Hsieh.
\newblock Large batch optimization for deep learning: Training bert in 76
  minutes.
\newblock {\em arXiv preprint arXiv:1904.00962}, 2019.

\bibitem{yu2022metaformer}
W.~Yu, C.~Si, P.~Zhou, M.~Luo, Y.~Zhou, J.~Feng, S.~Yan, and X.~Wang.
\newblock Metaformer baselines for vision.
\newblock {\em arXiv preprint arXiv:2210.13452}, 2022.

\bibitem{zhang2018mixup}
H.~Zhang, M.~Cisse, Y.~N. Dauphin, and D.~Lopez-Paz.
\newblock mixup: Beyond empirical risk minimization.
\newblock In {\em International Conference on Learning Representations}, 2018.

\bibitem{zhong2020random}
Z.~Zhong, L.~Zheng, G.~Kang, S.~Li, and Y.~Yang.
\newblock Random erasing data augmentation.
\newblock In {\em Proceedings of the AAAI conference on artificial
  intelligence}, volume~34, pages 13001--13008, 2020.

\end{thebibliography}

\newpage
\section{Appendix A: Training hyper-parameters}
\begin{table*}[!htbp]
\begin{center}
$
\begin{array}{l | c}
\toprule
\text{Configuration} & \text{AudioSet2M} \\ 
\hline \\
 \text{Optimizer} & \text{AdamW \cite{adamw}} \quad | \quad \text{LAMB \cite{you2019large}}^{\dag} \\ 
 \text{Optimizer momentum} & \beta_1=0.9, \beta_2=0.999 \\
 \text{Weight decay} & 0.05 \\
 \text{Base learning rate} & 4e-3 \\
 \text{Learning rate schedule} & \text{half-cycle cosine decay \cite{loshchilov2016sgdr}} \\
 \text{Gradient clipping} & \text{None} \\
 \text{Epochs} & 60 \\
 \text{Warm-up epochs} & 20 \\
 \text{Batch size} & 4096 \\
 \text{GPUs size} & 32 \\
 \text{Weighted sampling} & \text{False} \\
 \text{Drop path \cite{larsson2016fractalnet}} & 0.4 \\
 \text{Mixup \cite{zhang2018mixup}} & 0.8 \\
 \text{Multilabel} & \text{True} \\
  \text{Label smoothing \cite{szegedy2016rethinking}} & 0.1 \\
 \text{Loss Function} & \text{Binary Cross-Entropy} \\
 \text{Dataset Mean for Normalization} & -18.2696 \\
 \text{Dataset Std for Normalization} & 30.5735 \\
 \hline
 \multicolumn{2}{c}{\text{Spectrogram configuration}}\\\\
 \text{Number of fft} & 1024 \\
 \text{Hop length} & 320 \\
 \text{Power} & 2 \\
  \hline
 \multicolumn{2}{c}{\text{Mel scale configuration}}\\\\
 \text{Number of mels} & 128 \\
 \text{Sample rate} & 32 ~ 000 \\
 \text{Min frequency} & 50 \\
 \text{Max frequency} & 14 ~ 000 \\
 \text{Amplitude to dB} & \text{True} \\ 
  \hline
 \multicolumn{2}{c}{\text{Augmentations}}\\\\
 \text{PadOrTruncate} & 10 \times \text{sample rate} \\
 \text{RandomRoll} & [-\text{sample size}, \text{sample size}], \text{p} = 1 \\ 
 \text{SpeedPerturbation \cite{ko15_interspeech}} & \text{rates} = (0.5, 1.5),  \text{p} = 0.5 \\
 \text{RandomErasing \cite{zhong2020random}} & \text{p} = 0.25 \\

\end{array}
$
\end{center}
\caption{\textbf{Training hyper-parameters.}}
\label{tab:2}
\end{table*}
\newpage
\section{Appendix B: How to replace all model's DSC by DCLS ones ?}
\label{sec:appendixb}
\begin{lstlisting}[language=Python]
import copy
from torch import nn
from DCLS.construct.modules import Dcls2d

# Helper function that replaces all ".int." patterns
# by "[int]" in a character string
def replace_dots_brackets(name):
    name_split = name.split(".")
    name_split = [
        "[" + i + "]" if i.isdigit() else "." + i for i in name_split
    ]
    return "".join(name_split[:-1]), name_split[-1][1:]

# Helper function that replaces all the
# 2D depthwise separable convolution in
# a model by synchronized Dcls2d ones
def replace_depthwise_dcls(
    model, dilated_kernel_size=23, kernel_count=26, version="gauss"):
    in_channels, P, SIG = 0, None, None
    # Loop over all model modules
    for name, module in model.named_modules():
        # if the module is a depthwise separable Conv2d module
        if (isinstance(module, nn.Conv2d)
            and module.groups == module.in_channels == module.out_channels
            and module.kernel_size == (7, 7)
        ):
            name_eval, last_layer = replace_dots_brackets(name)
            dcls_conv = Dcls2d(
                module.in_channels,
                module.out_channels,
                kernel_count=kernel_count,
                stride=module.stride,
                dilated_kernel_size=dilated_kernel_size,
                padding=dilated_kernel_size // 2,
                groups=module.in_channels,
                version=version,
                bias=module.bias is not None,
            )
            nn.init.normal_(dcls_conv.weight, std=0.02)
            if module.bias is not None:
                nn.init.constant_(dcls_conv.bias, 0)

            # Synchronise positions and standard
            # deviations belonging to the same stage
            if in_channels < module.in_channels:
                in_channels = module.in_channels
                P, SIG = dcls_conv.P, dcls_conv.SIG

            dcls_conv.P, dcls_conv.SIG = P, SIG
            setattr(eval("model" + name_eval), last_layer, dcls_conv)
    return model

model = nn.Conv2d(96, 96, 7, padding=3, groups=96)
# Replace all the 2D depthwise separable convolutions
# in the model by synchronized Dcls2d ones.
model = replace_depthwise_dcls(
    copy.deepcopy(model),
    dilated_kernel_size=23,
    kernel_count=26,
    version="gauss",
)
print(model)
\end{lstlisting}
\end{document}